\documentclass{kapproc} 
\usepackage{procps} 
\usepackage[dvips]{graphicx}

\upperandlowercase

\kluwerbib 

\begin{document}

\articletitle[]{Magnetic Field Generation and electron acceleration 
in Collisionless Shocks}

\author{Christian B. Hededal\altaffilmark{1}, Jacob Trier Frederiksen\altaffilmark{2}, Troels Haugboelle\altaffilmark{1}, Aake Nordlund\altaffilmark{1}}

\altaffiltext{1}{Niels Bohr Institute for Astronomy, Physics, and Geophysics, Juliane Maries Vej 30, 2100 Koebenhavn Oe , Denmark}
\altaffiltext{2}{Stockholm Observatory, Roslagstullbacken 21, 106 91 Stockholm, Sweden}

\begin{abstract}
Using a three dimensional relativistic particle--in--cell code we have 
performed numerical experiments of plasma shells colliding at relativistic
velocities. Such scenarios are found in many astrophysical objects e.g. 
the relativistic outflow from gamma ray bursts, active galactic nuclei jets and
supernova remnants. We show how a Weibel like two--stream instability is
capable of generating small--scale magnetic filaments with strength up 
to percents of equipartition. Such field
topology is ideal for the generation of jitter radiation as opposed to
synchrotron radiation. We also explain
how the field generating mechanism involves acceleration of electrons to power law 
distributions ($N(\gamma)\propto{}\gamma^{-p}$) through a non--Fermi acceleration mechanism. The results
add to our understanding of collisionless shocks.
\end{abstract}


\section{Introduction}
Many astrophysical objects emit non--thermal radiation when expelled plasma 
interacts with the surrounding media. These objects include gamma 
ray bursts (GRBs) and their afterglows, the jets from active galactic nuclei, 
jets from quasars and supernova remnants. The non--thermal radiation is believed 
to be emitted in strong collisionless shocks. Despite the vast prevalence and wide 
astrophysical applicability, collisionless shocks remain poorly understood.
The observed non--thermal emission suggests that strong particle 
acceleration and magnetic 
field generation takes place in these shocks. It was suggested by 
\cite{1999ApJ...526..697M} that a Weibel--like two--stream instability is 
able to generate a strong magnetic field in the shock transition region. 
Recently, due to the increase in computer power, this has been verified with 
particle--in--cell (PIC) simulations (\cite{bib:astro-ph/0303360,2004ApJ...608L..13F,bib:astro-ph/0408558,bib:medv,2003ApJ...595..555N,bib:astro-ph/0409702,2003ApJ...596L.121S}).
Still, todays computer power does not allow us to fully resolve the
shock transition region. We can, however, explain the non-linear kinetic 
plasma dynamics that generate the electromagnetic fields 
needed for transmission of momentum between the colliding plasma populations.
Here, we report on 3D PIC simulations of the shock formation in the 
counter--streaming region of two colliding plasma shells.

\section{Numerical Experiments}
We have performed numerical experiments using a three dimensional relativistic
kinetic electromagnetic particle--in--cell code
The code works from first principles by solving 
the Lorentz force equation for the particles and the Maxwell's equations 
for the electromagnetic fields.

We let two electron-proton plasma populations 
(with a density difference of a factor of three) collide in the reference 
frame of the denser population. In this frame we continuously inject the 
less dense population with a bulk Lorentz factor, $\Gamma=15$ in the 
$z$-direction. 
The computational box consists of $125\times125\times2000$ gridzones or 
$37\times37\times600\Delta_e^3$ where $\Delta_e$ is the electron skin depth 
$c/\omega_e$. Using 16 particles pr. cell this adds up to almost $10^9$ particles. 
Both populations have a rest frame temperature corresponding to a thermal 
velocity $v_{th}=0.01c$. In order to be able to resolve both electron and ion
dynamics the ion-electron mass ratio is $M_i/m_e=16$. This is clearly 
a strong assumption but it allow us qualitatively to understand how 
particles with different mass affects the Weibel instability. In these 
experiments both populations are initially unmagnetized.

\section{Results}
\begin{figure}[h]
\includegraphics[width=\textwidth]{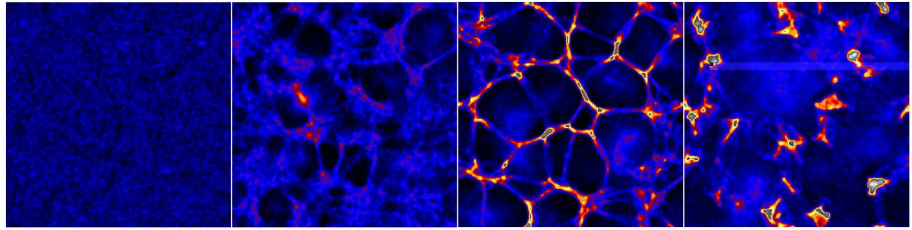}
\caption{This figure shows the generation of ion current filaments.
Here we see the jet head on. The four
slices show the ion current density at different depths of the shock at a fixed
time. The different depths are z=\{60, 100, 120, 160\} electron skin depths.}
\label{fig:scan}
\end{figure}

When the simulation runs and the plasma populations stream through each other,
we observe how the Weibel instability collects particles into current
filaments (\cite{1999ApJ...526..697M}). First the electrons go through the 
instability and then further downstream the electrons thermalize 
to one single population and the heavier ions goes through the instability. 
 The ion filaments are
 more robust than the preceding electron filaments since the thermalized 
 electron will Debye shield the ion filaments.
Even further downstream the current filaments acts as 2D macro
particles in the transverse plane and are themselves collected into larger 
filaments as explained by \cite{2004ApJ...608L..13F} and \cite{bib:medv}.
This behavior can be seen in fig. \ref{fig:scan}. 

Surrounding the current filaments is 
generated a strong magnetic field with $\epsilon_B\sim0.01$. 
Here, $\epsilon_B$ is the field generation efficiency i.e. the amount of 
total injected kinetic energy that goes into magnetic field energy. It is 
important to realize that $\epsilon_B$ is not a simple parameter since it
varies strongly along the flow-direction down  through the shock.

\begin{figure}[h]
\begin{center}
\includegraphics[width=5cm]{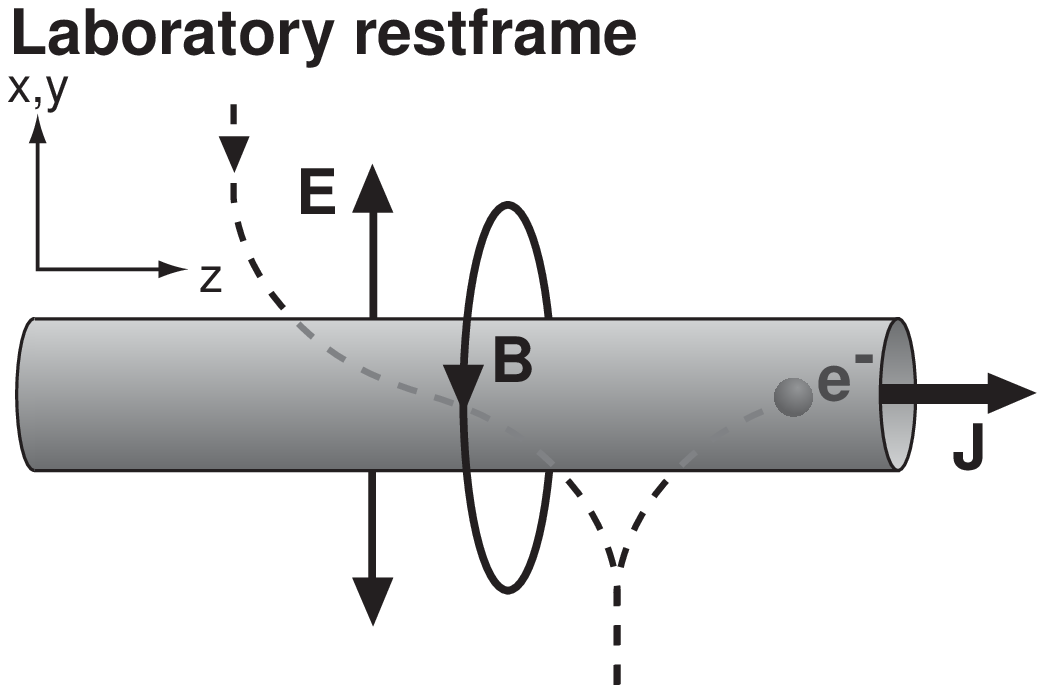}
\includegraphics[width=5.5cm]{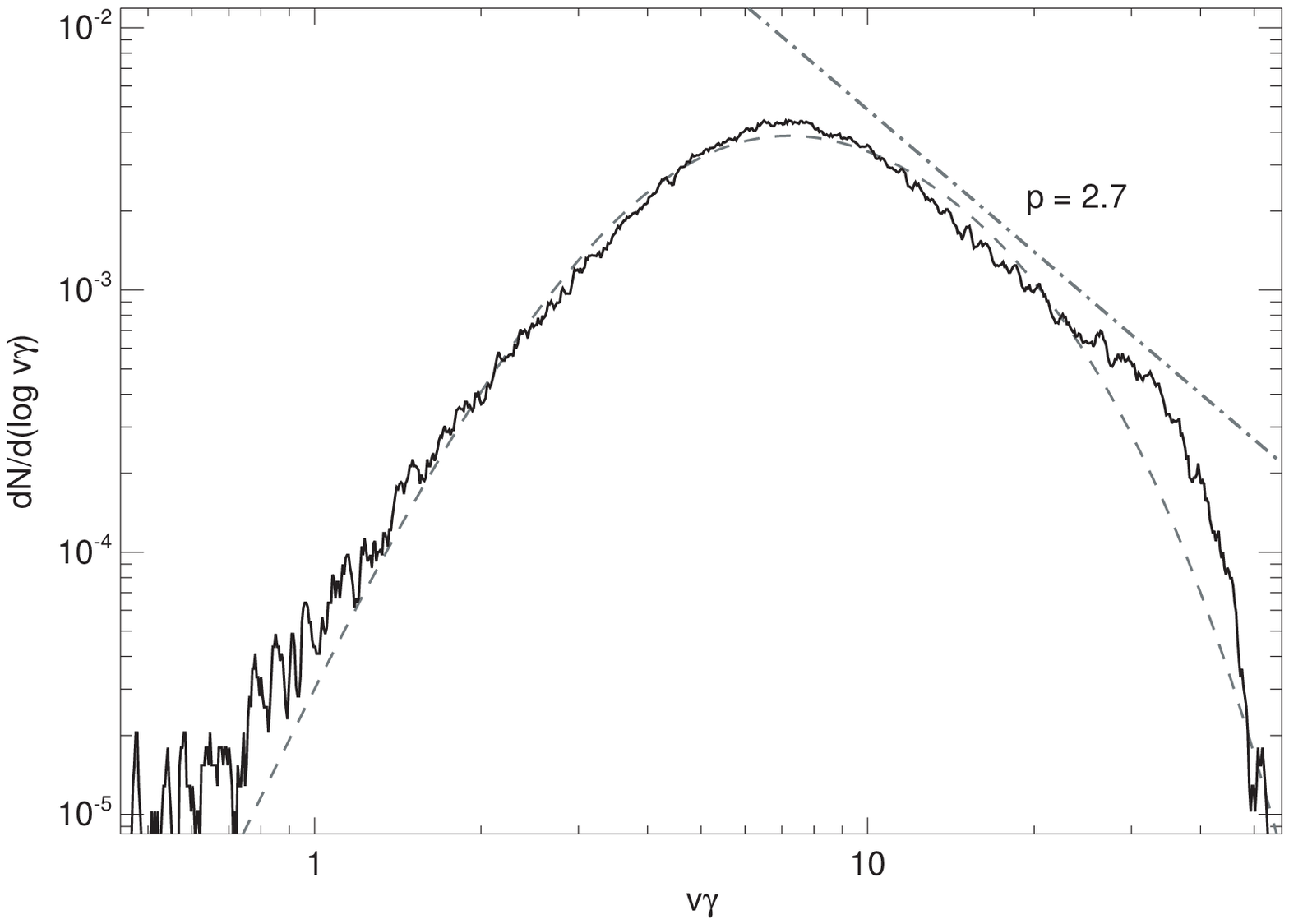}
\label{fig:hist}
\end{center}
\sidebyside
{\caption{An ion current channel surrounded by an electric -- and a magnetic field.
Electrons in the vicinity of the current channels are thus subject to a Lorentz force
with both an electric and magnetic component, effectively accelerating the
electrons.
}\label{fig:model}}
{\caption{The normalized electron four velocity distribution function downstream of
the shock. The dot--dashed line is a power law fit to the non--thermal high
energy tail, while the dashed curve is a Lorentz boosted thermal electron
population.
}\label{fig:distrib}}.
\end{figure}

Inside the Debye sphere, strong electron acceleration takes place. The 
electrical field that surrounds the ion current channel accelerates the 
electrons toward the filaments where they are deflected on the induced
magnetic field. The scenario is depicted in fig. \ref{fig:model}. It have
previously been shown that the ion filaments are generated in a self-similar
coalescence process (\cite{bib:medv}) which implies that a spatial Fourier 
decomposition exhibits power law behavior. As a result, the electrons are
accelerated to a power law distribution function (fig. 
\ref{fig:distrib}) as shown by \cite{bib:astro-ph/0408558}.

The electrons are trapped in the potential field of the ion channels and 
are repeatedly accelerated and decelerated. This means that energy losses 
due to escape are small, and that the electrons 
remain trapped long enough to have time to loose their energy via a 
combination of bremsstrahlung and synchrotron or jitter radiation. The fact
that the electrons cannot escape the ion filaments without being decelerated 
also implies that they are are not available for recursive acceleration as
suggested in Fermi acceleration (\cite{bib:astro-ph/0408558}).
\section{Summary}
We have performed numerical experiments of relativistic collisionless plasma
shocks using a self-consistent three dimensional particle--in--cell code.
We find that the Weibel instability is capable of generating
turbulent magnetic fields with a strength up to percents of equipartition.
The magnetic field is induced around ion current filaments. These filaments
also accelerates electrons to power law distributions.
The suggested acceleration scenario does not rule out ion Fermi acceleration
but might overcome some of the 
problems pointed out by \cite{2004ApJ...613..460B} regarding the
apparent contradiction between standard Fermi acceleration of electrons
and spectral observations of GRBs.

The microphysics in the field generation and particle acceleration described
here is clearly beyond the reach of the magneto hydrodynamic approximation.
A parameter study utilizing a PIC code working from first principles
is necessary to fully understand the interdependence between the relative
bulk Lorentz factors of the colliding plasma shells, the power law index of the
non--thermal electron population, $\epsilon_B$ and in a broader sense the
detailed evolution and structure in collisionless shocks.
\bibliographystyle{kapalike}
\chapbblname{hededal}
\chapbibliography{hededal}

\end{document}